\documentclass[fleqn,usenatbib]{mnras}
\usepackage{newtxtext,newtxmath}
\usepackage[T1]{fontenc}

\usepackage{graphicx}
\usepackage{multicol}
\usepackage{subcaption}
\usepackage{amsmath}
\usepackage{mathrsfs}
\usepackage{amsfonts}
\usepackage{ulem}

\DeclareRobustCommand{\VAN}[3]{#2}
\let\VANthebibliography\thebibliography
\def\thebibliography{\DeclareRobustCommand{\VAN}[3]{##3}\VANthebibliography}

\title[Measuring mass transfer of AM CVns with GWs]{Measuring mass transfer of AM CVn binaries with a space-based gravitational wave detector}

\author[Zijian Wang et al.]{
Zijian Wang,$^{1,2}$
Zhoujian Cao,$^{1,2,3}$\thanks{E-mail: zjcao@amt.ac.cn}
Xian-Fei Zhang$^{1,2}$
\\
$^{1}$Institute for Frontiers in Astronomy and Astrophysics, Beijing Normal University, Beijing 102206, China\\
$^{2}$Department of Astronomy, Beijing Normal University, Beijing 100875, China\\
$^{3}$School of Fundamental Physics and Mathematical Sciences, Hangzhou Institute for Advanced Study, UCAS, Hangzhou 310024, China
}

\date{Accepted XXX. Received YYY; in original form ZZZ}

\pubyear{2023}

\begin{document}
\label{firstpage}
\pagerange{\pageref{firstpage}--\pageref{lastpage}}
\maketitle

\begin{abstract}
The formation mechanism of AM CVn binary has not been well understood yet. Accurate measurements of the mass transfer rate can help to determine the formation mechanism. But unfortunately such observation by electromagnetic means is quite challenging. One possible formation channel of AM CVn binary is a semi-detached white dwarf binary. Such system emits strong gravitational wave radiation which could be measured by the future space-based detectors. We can simultaneously extract the mass transfer rate and the orbital period from the gravitational wave signal. We employ a post-Keplerian waveform model of gravitational wave and carry out a Fisher analysis to estimate the measurement accuracy of mass transfer rate through gravitational wave detection. Special attention is paid to the observed sources in Gaia Data Release 2. We found that we can accurately measure the mass transfer rate for those systems. Comparing to electromagnetic observations, gravitational wave detection improves the accuracy more than two orders of magnitude. Our results imply that the gravitational wave detection will help much in understanding the formation mechanism of AM CVn binaries.
\end{abstract}


\section{Introduction}
LIGO-VIRGO-KAGRA Collaboration has successfully conducted observation runs 1-3. About 100 gravitational wave (GW) events have been reported. The observation run 4 will start in May 2023 soon. The GW astronomy develops rapidly (\cite{LV1,LV2,LV3,LV4,LV5}). The ground-based detectors aim at frequencies above 10 Hz. The space-based gravitational wave interferometers will expand the GW spectrum to lower frequency band ($10^{-4}\sim 1$ Hz). The undergoing space-based detectors include Laser Interferometer Space Antenna (LISA) (\cite{LISA}), TianQin (\cite{tianqin}) and Taiji (\cite{2020ResPh..1602918L,taiji}).

The ultracompact binaries (UCBs) with orbital periods ranging from several minutes to hours are important sources for space-based detectors. Such UCBs include white dwarf binaries (\cite{brown2020,burdge2020}), ultracompact X-Ray binaries (UCXBs) (\cite{chen2020}) and so forth. Some UCB systems could be observed by both the electromagnetic (EM) means and the GW means. The LISA verification binaries (\cite{kupfer2018}) belongs to this type including detached white dwarfs, hot subdwarf binaries and AM Canum Venaticorum (AM CVn) systems.

AM CVns are semidetached binaries undergoing helium mass transfer. Their orbital periods range from 5 to 65 minutes (\cite{amcvn}). These systems are ideal laboratories to test tidal interaction models, detonation mechanism, accretion theories (\cite{piro2019, ashok2003, bildsten2007, kotko2012}), etc. So far, the formation mechanism of AM CVn has remained undetermined. There are three possible channels corresponding donor evolution of He-core white dwarfs (\cite{1967, deloye2007}), nondegenerate He stars \cite{1986,1987,yungelson2008} and evolved hydrogen-rich CVs (\cite{tutukov1985,tutukov1987,2003,2015}), respectively. In each channel, the orbit evolution is initially dominated by gravitational wave radiation (GWR). GWR shrink the orbit rapidly until the Roche lobe overflow (RLOF) begins. Subsequently the mass transfer rate (MTR) rises to the peak and tends to dominate the evolution which turns down the orbital frequency increment and eventually makes it decrease. Precise measurement of the donor's mass transfer rate at a certain orbital period would help us to understand the AM CVn formation mechanism because different channels follow different tracks in $P_{\mathrm{orb}}-\dot{M}_{\mathrm{donor}}$ plane. Nowadays, MTR could be measured through observing bright spots in AM CVn systems. For example, \cite{ramsay2018} has determined the mass transfer rate for 15 AM CVn stars using Gaia Data Release 2 (Gaia DR2). Furthermore, a large amount of investigations on the Helium transfer have been carried out for double He WD channel (\cite{zhangxf2018}), CO+He WD channel (\cite{wong2021,chen2022}) and hot subdwarf B+WD scenario (\cite{Bauer_2021}).

LISA verification binaries could be detected with SNR$>$10 based on four year GW observation (\cite{kupfer2018}). Galactic binary WDs have high SNR and are usually regarded as continuous GW sources. Four year data and appropriate waveform models could make accurate parameter estimations of these sources. For example, LISA could measure the frequency $f$ and its time derivative $\dot{f}$ (\cite{shah2012}). In semi-detached binaries, $\dot{f}$ is affected by GWR, tidal forces and mass transfer (\cite{kremer2015}). Therefore, we could extract these parameters including mass transfer rate through matched filtering of GW waveform.

In this paper, we focus on measuring mass transfer in AM CVn systems especially in the white dwarf channel. In next section we express the chirp $\dot{f}$ in terms of frequency $f$ and mass transfer rate $\dot{M}_{\mathrm{donor}}$. Subsequently we carry out a Fisher information analysis for parameter estimation and present the measurability of mass transfer rate in different conditions in Sec.~\ref{sec3}. In Sec.~\ref{sec4} we forecast the GW measurement performance in $P_{\mathrm{orb}}-\dot{M}_{\mathrm{donor}}$ plane for several binaries in Gaia DR2 data. At last we conclude our paper in Sec.~\ref{sec5}.

\section{Waveform model for mass transfer measurement}\label{sec2}
We adopt the parameterized post-Keplerian gravitational waveform model for the white dwarfs in AM CVn binaries. Within the post-Keplerian waveform model, the orbital elements and their adiabatic variations involve the gravity theories and detailed stellar dynamics (\cite{damour1992}). This approximation is valid especially when the binary seperation is large and consequently the gravitational interaction is weak. Therefore, we could always describe the motion of the binary as a post-Keplerian expansion. The related general gravitational waveform can be expressed as (\cite{willems2008,2004PhRvD..69h2005B,2023GReGr..55...76L}):
\begin{eqnarray}
h^+(t) &=&\sum_{n=1}^\infty A\{(1+\cos^2\iota)a_n(e)\cos[n\phi(t)]\cos(2\gamma)\nonumber\\
&-&(1+\cos^2\iota)b_n(e)\sin[n\phi(t)]\sin(2\gamma)\nonumber\\
&+&(\sin^2\iota)c_n(e)\cos[n\phi(t)]\},\label{hptotal}\\
h^\times(t) &=&-\sum_{n=1}^\infty 2A\cos\iota\{b_n(e)\sin[n\phi(t)]cos(2\gamma)\nonumber\\
&+&a_n(e)\cos[n\phi(t)]\sin(2\gamma)\}\label{hctotal}
\end{eqnarray}
with
\begin{eqnarray}
a_n(e)&=&n[J_{n-2}(ne)-2eJ_{n-1}(ne)+\dfrac{2}{n}J_n(ne)\nonumber\\
&+&2eJ_{n+1}(ne)-J_{n+2}(ne)],\\
b_n(e)&=&n\sqrt{1-e^2}[J_{n-2}(ne)-2J_n(ne)\nonumber\\
&+& J_{n+2}(ne)],\\
c_n(e)&=&J_n(ne),
\end{eqnarray}
where $J_n(z)$ is the $n$th order Bessel function of the first kind and $\iota, ~e,~\gamma$ are inclination angle, orbit eccentricity and periastron direction, respectively. $\phi(t)$ is the Doppler shifted orbital phase
\begin{equation}
\phi(t)=\phi_0+\pi f t,
\end{equation}
where $f$ is the frequency of the gravitational waves. In order to impose constraints on the number of parameters, we expand GW frequency to the third order of time derivative:
\begin{equation}
f=f_0+\dot{f}_0t+\dfrac{1}{2}\ddot{f}_0t^2+\dfrac{1}{6}\dddot{f}_0t^3.
\end{equation}
In the post-Keplerian waveform model, time dependent variables change slowly over time. One could expand these variables adiabatically. The three orbital elements can be written as
\begin{eqnarray}
\iota&=&\iota_0+\dot{\iota}_0t,\\
e&=&e_0+\dot{e}_0t,\\
\gamma&=&\gamma_0+\dot{\gamma}_0t.
\end{eqnarray}

For WD binaries which may be affected by the mass transfer, we select circular orbit with $e_0=\dot{e}_0=0$ and thus $\gamma_0=\dot{\gamma}_0=0$. These assumptions are reasonable because the progenitors of double white dwarfs may experience several mass transfer phases and tidal forces would circularize the orbit (\cite{willems2007}). Once the mass transfer starts, tidal forces and gravitational radiation will keep the orbit circular (\cite{kremer2015}). Therefore, Eqn.~\eqref{hptotal} and Eqn.~\eqref{hctotal} can be simplified as
\begin{eqnarray}
h_+(t)&=&2A_0(1+\cos^2\iota)\cos[2\phi(t)],\\
h_\times(t)&=&-4A_0\cos\iota\sin[2\phi(t)],
\end{eqnarray}
with the amplitude
\begin{equation}
A_0=\dfrac{(GM_c)^{5/3}}{c^4D_L}(\pi f)^{2/3},
\end{equation}
where $M_c$ is chirp mass and $D_L$ is luminosity distance.

Now we take orbital evolution into consideration and assume the loss of angular momentum is only due to the gravitational radiation. We force both stars to remain non-rotating and thus there is no L-S coupling (also called spin-orbit coupling or Russell-Saunders coupling) throughout. The reason why this assumption is selected is that tides could significantly synchronize the two components and transfer the accreted spin angular momentum back to the orbit. Shorter synchronization timescales would help double white dwarfs (DWD) survive mass transfer and would be a better approximation for systems with short orbital periods. (\cite{marsh2004,fuller_lai_2012,fuller_lai_2014}). Observations of GP Com and V396 Hya also exclude the fast-rotating accretor models (\cite{kupfer2016}). Ignoring L-S coupling means that the direction of the orbital momentum is constant and thus the inclination is invariant over time. Therefore, $\dot{\iota}_0$ vanishes.

When one star of the binary expands beyond its Roche lobe, the mass transfer starts and the system becomes semidetached. The star with its Roche lobe filled is called the secondary star. It is also called the donor star. Its mass is traditionally denoted as $M_2$. The donor's companion is called the primary star or accretor whose mass is denoted as $M_1$ (\cite{carroll2017}). On the onset of mass transfer, the donor's atmospheric gases escape through the inner Lagrangian point $L_1$ to the accretor. We assume the mass transfer is conservative which means $\dot{M}=0$ with $M\equiv M_1+M_2$ and otherwise more parameters would be introduced, such as mass transfer rate induced by stellar winds and efficiency of accretion (\cite{1997A&A...327..620S}). In this work, we mainly investigate the measurability of mass transfer in DWDs and the interesting analysis related to detailed mechanisms can be done then. Accordingly we have $\dot{J}_{\mathrm{mass~loss}}=0$ for the binary system and the orbital angular momentum loss is only due to the emission of gravitational waves. Hence, the total change of the orbital angular momentum is given by
\begin{equation}
\dot{J}_{\rm orb}=\dot{J}_{\rm GR}.
\end{equation}
The orbital angular momentum is written as
\begin{equation}
J_{\rm orb}=\sqrt{\dfrac{Ga}{M}}M_1M_2\label{Jorb}
\end{equation}
and the change of the orbital angular momentum induced by GWR is given by \cite{landau}
\begin{equation}
\dot{J}_{\rm GR}=-\dfrac{32G^3}{5c^5}\dfrac{M_1M_2M}{a^4}J_{\rm orb}.\label{dJGR}
\end{equation}
From Eqn.~\eqref{Jorb}, the change of the orbital angular momentum follows
\begin{equation}
\dfrac{\dot{J}_{\rm orb}}{J_{\rm orb}}=(1-q)\dfrac{\dot{M}_2}{M_2}+\dfrac{\dot{a}}{2a},\label{dJorb}
\end{equation}
where $q\equiv\dfrac{M_2}{M_1}$. The Eqn.~\eqref{dJGR} and Eqn.~\eqref{dJorb} could be combined to obtain the change of semi-major axis of the binary
\begin{equation}
\dfrac{\dot{a}}{2a}=\dfrac{\dot{J}_{\rm GR}}{J_{\rm orb}}-(1-q)\dfrac{\dot{M}_2}{M_2}.\label{da}
\end{equation}
According to the Kepler's third law, the frequency of the orbit is
\begin{equation}
f_{\rm orb}=\left(\dfrac{GM}{4\pi^2a^3}\right)^{1/2}
\end{equation}
and the GW frequency is
\begin{equation}
f=2f_{\rm orb}.
\end{equation}
The time derivative of $f$ which is called the chirp is given by
\begin{equation}
\dot{f}=-3\dfrac{\dot{a}}{2a}f\label{df}.
\end{equation}
Combining Eqn.~\eqref{df} with Eqn.~\eqref{da} and replacing $M_1,~M_2$ with $\eta,~M_c$, the change of the frequency could be written as
\begin{equation}
\dot{f}=\dot{f}_{\rm 0PN}(1+\Delta_{\rm 1PN}f^{2/3})+\Delta_{\rm MT}f.\label{dftotal}
\end{equation}
with
\begin{eqnarray}
\dot{f}_{\rm 0PN}&=&\dfrac{96}{5}\dfrac{(GM_c)^{5/3}\pi^{8/3}}{c^5}f^{11/3},\\
\Delta_{\rm 1PN}&=&-\left(\dfrac{743}{336}+\dfrac{11}{4}\eta\right)\left(\dfrac{\pi M_c G}{\eta^{3/5}c^3}\right)^{2/3},\\
\Delta_{\rm MT}&=&\dfrac{6\sqrt{1-\eta}}{M_c\eta^{2/5}}\dot{M}_2,\label{deltaMT}
\end{eqnarray}
where $\eta=\dfrac{M_1 M_2}{M^2}$ is the symmetric mass ratio and $M_c$ is the so-called chirp mass $M_c\equiv(M_1M_2)^{3/5}M^{-1/5}$. The first term of Eqn. \eqref{dftotal} is the leading order in the post Newtonian expansion and the second term is the next-to-leading order which depends on mass ratio as well (\cite{arun2005}). The third term is due to mass transfer and mass transfer rate is set to be negative, $\dot{M}_2<0$. Similarly, the second and the third time derivatives of GW frequency $f$ are given by
\begin{eqnarray}
\ddot{f}&=&\dfrac{11}{3}\dfrac{\dot{f}_{\rm 0PN}^2}{f}\left(1+\dfrac{24}{11}\Delta_{\rm 1PN}f^{2/3}\right)+\left.\dfrac{14}{3}\Delta_{\rm MT}\dot{f}_{\rm 0PN}\right(1\nonumber\\
&+&\left.\dfrac{8}{7}\Delta_{\rm 1PN}f^{2/3}\right)+\Delta_{\rm MT}^2f,\\
\dddot{f}&=&\dfrac{209}{9}\dfrac{\dot{f}_{\rm 0PN}^3}{f^2}\left(1+\dfrac{713}{209}\Delta_{\rm 1PN}f^{2/3}\right)+\dfrac{121}{3}\dfrac{\Delta_{\rm MT}\dot{f}_{\rm 0PN}^2}{f}\nonumber\\
&\cdot&\left(1+\dfrac{366}{363}\Delta_{\rm 1PN}f^{2/3}\right)+\left.\dfrac{163}{9}\Delta_{\rm MT}^2\dot{f}_{\rm 0PN}\right(1\nonumber\\
&+&\left.\dfrac{217}{164}\Delta_{\rm 1PN}f^{2/3}\right)+\Delta_{\rm MT}^3f.
\end{eqnarray}

In the detector frame the waveform can be expressed as (\cite{willems2008})
\begin{equation}
h=\dfrac{\sqrt{3}}{2}[F_+h^++F_\times h^\times]
\end{equation}
where $F_+,~F_\times$ are pattern functions of the detector
\begin{eqnarray}
F_+ &=& \dfrac{1}{2}(1+\sin^2\theta)\cos2\phi\cos2\psi\nonumber\\
&-&\sin\theta\sin2\phi\sin2\psi,\\
F_\times &=& \dfrac{1}{2}(1+\sin^2\theta)\cos2\phi\sin2\psi\nonumber\\
&+&\sin\theta\sin2\phi\cos2\psi.
\end{eqnarray}
In the current work, we focus on the effects of the mass transfer and thus we can ignore the source localization and the wave polarization. Therefore we fix $\phi=\psi=0$ and $\theta=\pi/2$ and thus the pattern functions are equivalent to sky-averaging pattern functions when source position coincides with the specific direction. We assume that intrinsic parameters are independently of the precise location of the source as well as the orientation of the spacecraft and thereby we average over the pattern functions which could be written as
\begin{equation}
F_+=1,~~F_\times=0.
\end{equation}
Hence, we only consider 6 intrinsic parameters($A,~f_0,~q,~M_c,~\dot{M}_2,\\~\phi_0$). In order to avoid parameter degeneracy, we use $A$ to replace $2A_0(1+\cos^2\iota)$. In the following we can use matched filtering to estimate the waveform parameter $\dot{M}_2$.

\section{Parameter Estimation}\label{sec3}
We use Fisher information analysis to estimate the measurement accuracy of the aforementioned waveform parameters. Fisher information techniques have been used to estimate measurement accuracy in many works (see e.g. \cite{cutler},  \cite{shah2012}, \cite{shah2014} and \cite{wolz2021}). For large signal to noise ratio $S/N$, the parameter-estimation errors have the Gaussian probability distribution (\cite{cutler}) and then the Fisher information matrix (FIM) could provide the GW parameter measurement errors based on appropriate gravitational wave templates. The measurement uncertainty of the parameters $\lambda^i$ at $1\sigma$ confidence is given by the inverse of the FIM $\Gamma_{ij}$
\begin{equation}
\Delta\lambda^i=\sqrt{\Gamma^{-1}_{ii}},
\end{equation}
where $\Gamma_{ij}$ is defined by
\begin{equation}
\Gamma_{ij}=\left(\left.\dfrac{\partial h}{\partial\lambda^i}\right|\dfrac{\partial h}{\partial\lambda^j}\right).
\end{equation}
The inner product $(~|~)$ is defined as (\cite{finn})
\begin{equation}
(a(t)|b(t))=4\Re\int_0^\infty\dfrac{\tilde{a}^*(f)\tilde{b}(f)}{S_n(f)}df,\label{intf}
\end{equation}
where the asterisk denotes complex conjugation, the tilde denotes the Fourier transformation and $S_n$ is the spectral noise density of the detector. Since within the time of observation the signal of the white dwarf binaries could be treated as monochromatic, we make an approximation of $S_n(f)\approx S_n(f_0)$ and Eqn.~\eqref{intf} could be turned into time integral (\cite{wolz2021})
\begin{equation}
(a|b)=\dfrac{2}{S_n(f_0)}\int_0^{T_{\rm obs}}a(t)b(t)dt.
\end{equation}
The sensitivity curve of LISA is given by \cite{LISASn}:
\begin{eqnarray}
S_n(f) &=& \dfrac{10}{3L^2}\left(P_{\mathrm{OMS}}+2(1+\cos^2(f/f_*))\dfrac{P_{\mathrm{acc}}}{(2\pi f)^4}\right)\nonumber\\
&\times&\left(1+\frac{6}{10}\left(\dfrac{f}{f_*}\right)^2\right),\\
f_* &=& c/(2\pi L),\\
P_{\mathrm{OMS}} &=&(1.5\times10^{-11}\mathrm{m})^2\mathrm{Hz}^{-1},\\
P_{\mathrm{acc}} &=&\left(3\times10^{-15}\mathrm{m~s^{-2}}\right)^2\nonumber\\
&\times&\left[1+\left(\dfrac{4\times10^{-4}\mathrm{Hz}}{f}\right)^2\right]\mathrm{Hz}^{-1},\\
L &=& 2.5\times10^9\mathrm{m}.
\end{eqnarray}
The observation time is chosen to be maximal running time of the satellites which is $T_{\mathrm{obs}}=10$ years.

\section{Mass transfer rate estimation}\label{sec4}
In the following we firstly investigate the detectability of AM CVn's mass transfer rate. After that we pay special attention to the semi-detached systems. The known verification binaries of LISA belongs to this type and they may undergo mass transfer phase. Although GW observation data on them have not been obtained so far, mass transfer rates in white dwarfs binary simulations and optical observation results are used for parameter estimation.

\subsection{Detectability}
The WD scenario of AM CVn consists of a degenerate, low mass He WD with the mass ranging from 0.1 $M_\odot$ to 0.2 $M_\odot$ and a 0.6-1.0 $M_\odot$ carbon-oxygen WD (\cite{wong2021}). We present the GW measurement results of MTRs in a white dwarf binary with individual masses of $m_1=0.75M_\odot$, $m_2=0.15M_\odot$. In addition we set observation frequency $f_0=5\times10^{-3}$ Hz, the orbital inclination $\iota=\dfrac{\pi}{2}$ and the initial phase $\phi_0=0$ as fiducial values. We vary the mass transfer rate $\dot{M}_2$ and the luminosity distance $D_L$ in order to discuss the measurement accuracy of $\dot{M}_2$ with LISA.

\begin{figure}
\centering
\includegraphics[width=0.5\textwidth]{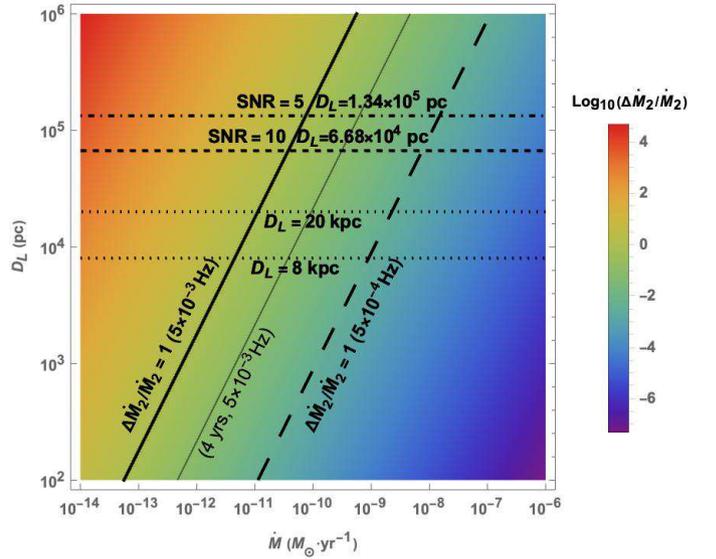}
\caption{Measurement error of the mass transfer rate $\Delta\dot{M}_2$ at fiducial frequency $f_0=5\times10^{-3}$ Hz with a 10-year observation of LISA. The detection threshold is constrained by the black solid line ($\Delta\dot{M}_2/\dot{M}_2$) and the dot dashed line (SNR). The black solid line represents $\Delta\dot{M}_2/\dot{M}_2=1$ and the dot dashed line corresponds to the SNR under 5 corresponding to the maximum detectable range $1.34\times10^5$ pc. The black dashed line indicates that the SNR $= 10$ at $6.68\times10^4$ pc. Double white dwarfs in the galactic center and the population edge have distances of 8 kpc and 20 kpc which are plotted as dotted lines. For comparison, we present the threshold for WD binaries at observation frequency $f_0=5\times10^{-4}$ Hz which is corresponding to the orbital period, $P_{\rm orb}=65$ min. When the observation time is set to be 4 years, the detection threshold at the frequency $f_0=5\times10^{-3}$ Hz is depicted to be the fine line.}\label{errors}
\end{figure}

Neglecting the dynamical evolution of the stellars, the measurement error at 1 $\sigma$ is a function of the MTR of the system and the luminosity distance. As shown in Fig.~\ref{errors}, we vary the MTR from $10^{-14}~M_\odot/\mathrm{yr}$ to $10^{-6}~M_\odot/\mathrm{yr}$ and the luminosity distance $D_L$ among the range $10^2\sim10^6$ pc in the $\dot{M_2}-D_L$ parameter plane. The detectability could be described by two indicators including the SNR and the measurement errors. For a given detector, the SNR is directly associated with the signal amplitudes which are inversely proportional to the luminosity distance. At luminosity distance $D_L=6.68\times10^4$ pc, the SNR is about $10$. When $D_L>1.34\times10^5$pc, the SNR would decrease below about 5. We select $10^5$ pc as the maximum detectable range. Denoting the measurement error of MTR as $\Delta\dot{M}_2$, we regard $\Delta\dot{M}_2/\dot{M}_2<1$ as measurable. $\Delta\dot{M}_2/\dot{M}_2=1$ corresponds to the black solid line plotted in Fig.~\ref{errors}. For a WD binary, if time derivative of frequency is negative, or to say, the orbital period is increasing, the system is dominated by astrophysical effects other than the GWR. According to the currently known AM CVn systems ($P_{\rm orb}\sim5-65$ min) with periods increasing, the source distances are distributed between several hundred and two thousand parsecs and the MTR ranges from $10^{-12}M_\odot/\mathrm{yr}$ to $10^{-7}M_\odot/\mathrm{yr}$ (see \cite{ramsay2018} for more information). When the distance is around $10^3$ pc, we could measure the MTR accurately if only $\dot{M}_2>5\times10^{-13}M_\odot/\mathrm{yr}$. When the He WD donor fills its Roche lobe, the MTR rapidly peaks to $10^{-7}\sim10^{-8}M_\odot/\mathrm{yr}$ within $10^6$ yrs and then decreases as $t^{-1.3}$ (\cite{bildsten2006}). The other two important distances are 8 kpc and 20 kpc which are distances of sources in galactic center and edge of the population estimated by \cite{kremer2017}. At the galactic boundary, the MTR could be measured precisely if $\dot{M}_2>6\times10^{-12}M_\odot/\mathrm{yr}$ and at the center, the threshold is $3\times10^{-12}M_\odot/\mathrm{yr}$. Therefore, the GW measurement abilities of the MTR in WD channel of AM CVn systems can almost cover the whole stage from the onset of mass transfer to the end of evolution. Additionally, we found that the measurability depends on initial frequency $f_0$. For most galactic AM CVn systems with long periods, such as $P_{\rm orb}\sim65$ min corresponding to $f_0\sim5\times10^{-4}$ Hz, the mass transfer rate could still be measurable when it's large enough to exceed $10^{-10}M_\odot/\rm yr$. Since the normal mission lifetime is 4 years, here we depicts the measurement threshold to be the fine solid line in Fig. \ref{errors}. Results show that at the distance of $10^3$ pc, we could measure the MTR with high accuracy if $\dot{M}_2>4\times10^{-12}M_\odot/\rm yr$ within a 4-year observation. And at the boundary and center of the Milky Way galaxy, the threshold are $8\times10^{-11}M_\odot/\rm yr$ and $3    \times10^{-11}M_\odot/\rm yr$, respectively.

\subsection{Applications}
\begin{figure*}
\centering
\includegraphics[width=0.95\textwidth]{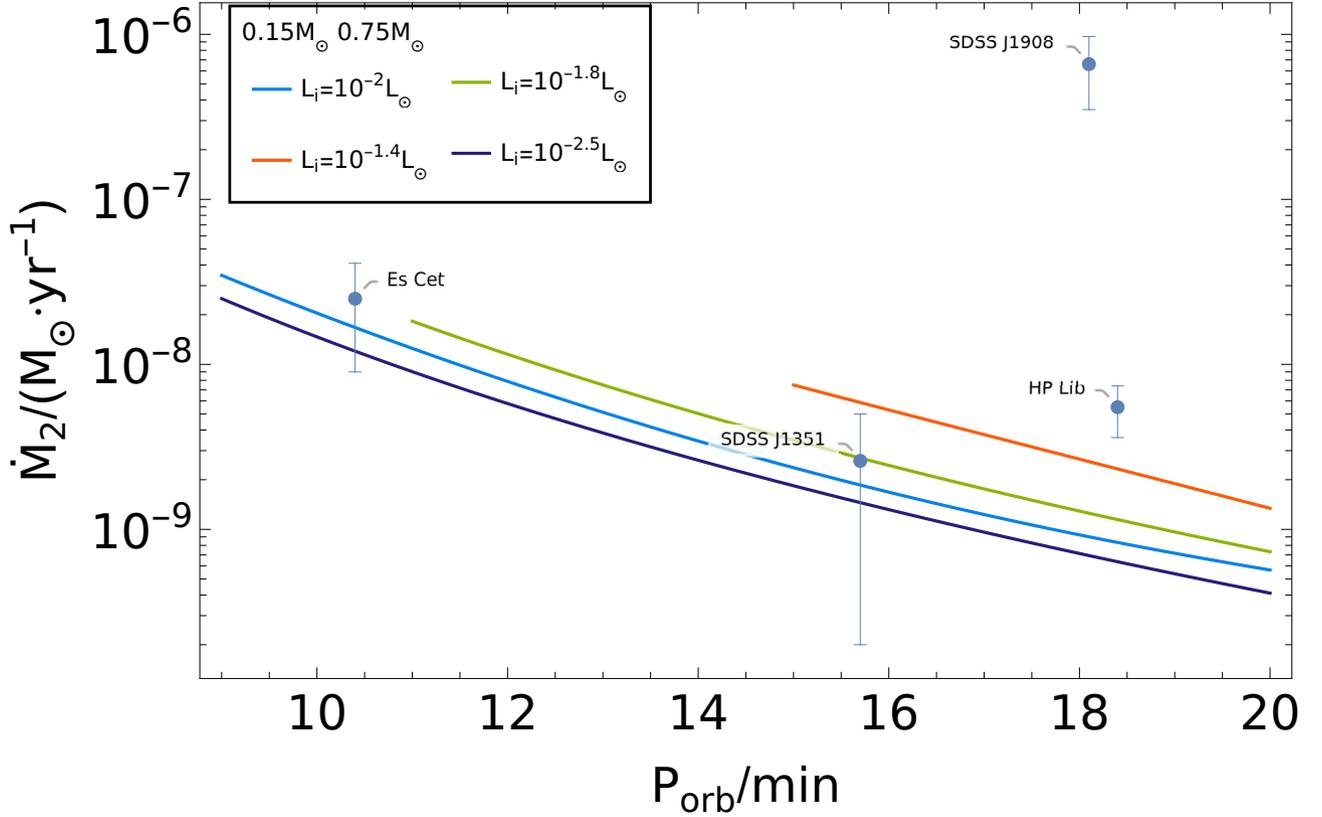}
\caption{Optical observations of mass transfer rate for AM CVn systems reported by Gaia DR2 among the LISA verification binaries (\protect\cite{ramsay2018}). These systems are ES Cet, SDSS J1351-0643, SDSS J1908+3940 and HP Lib with respectively periods 10.4, 15.7, 18.1, 18.4 minute and distances 1584, 1317, 1044, 276 pc. The colored lines are the evolution of double white dwarfs of $m_1=0.75M_\odot$, $m_2=0.15M_\odot$ with different initial luminosities of the donor. Higher luminosity means less internal material degeneracy and larger radius which induces mass transfer start at longer periods.}\label{opt}
\end{figure*}
Next we will discuss the applications to realistic astrophysical problems. In current optical observations, reliable values of mass transfer rate require a wide-band spectrum. Using Gaia DR2, \cite{ramsay2018} have determined the mass transfer rate for 15 AM CVn stars. Four of these sources are LISA verification binaries which are expected to emit detectable GW signals (\cite{kupfer2018}), as shown in Fig.~\ref{opt}. They are ES Cet, SDSS J1351-0643, SDSS J1908+3940 and HP Lib respectively with periods 10.4, 15.7, 18.1, 18.4 minute and distances 1584, 1317, 1044, 276 pc. The mass transfer rate of these AM CVn systems is important to distinguish three possible formation channels. Different channel follows different tracks in the $P_{\mathrm{orb}}-\dot{M}_{\mathrm{donor}}$ plane. This is because the mass transfer rate $\dot{M}_{\mathrm{donor}}$ depends on the donor's mass-radius relation which is determined by its internal material degeneracy and thermal evolution (\cite{deloye2007}).

The over-fill of the Roche lobe is defined as (\cite{marsh2004}):
\begin{equation}
\Delta=R_{\mathrm{donor}}-R_L,
\end{equation}
where $R_L$ is the radius of the Roche lobe. \cite{eggleton1983}'s approximation for the Roche lobe radius gives the expression as
\begin{equation}
R_L=a\frac{0.49q^{2/3}}{0.6q^{2/3}+\ln(1+q^{1/3})},
\end{equation}
where $a$ is the semi-major axis. When $R_L$ is obtained, the binary period could be determined through $\Delta = 0$.

\begin{figure*}
\centering
\includegraphics[width=0.95\textwidth]{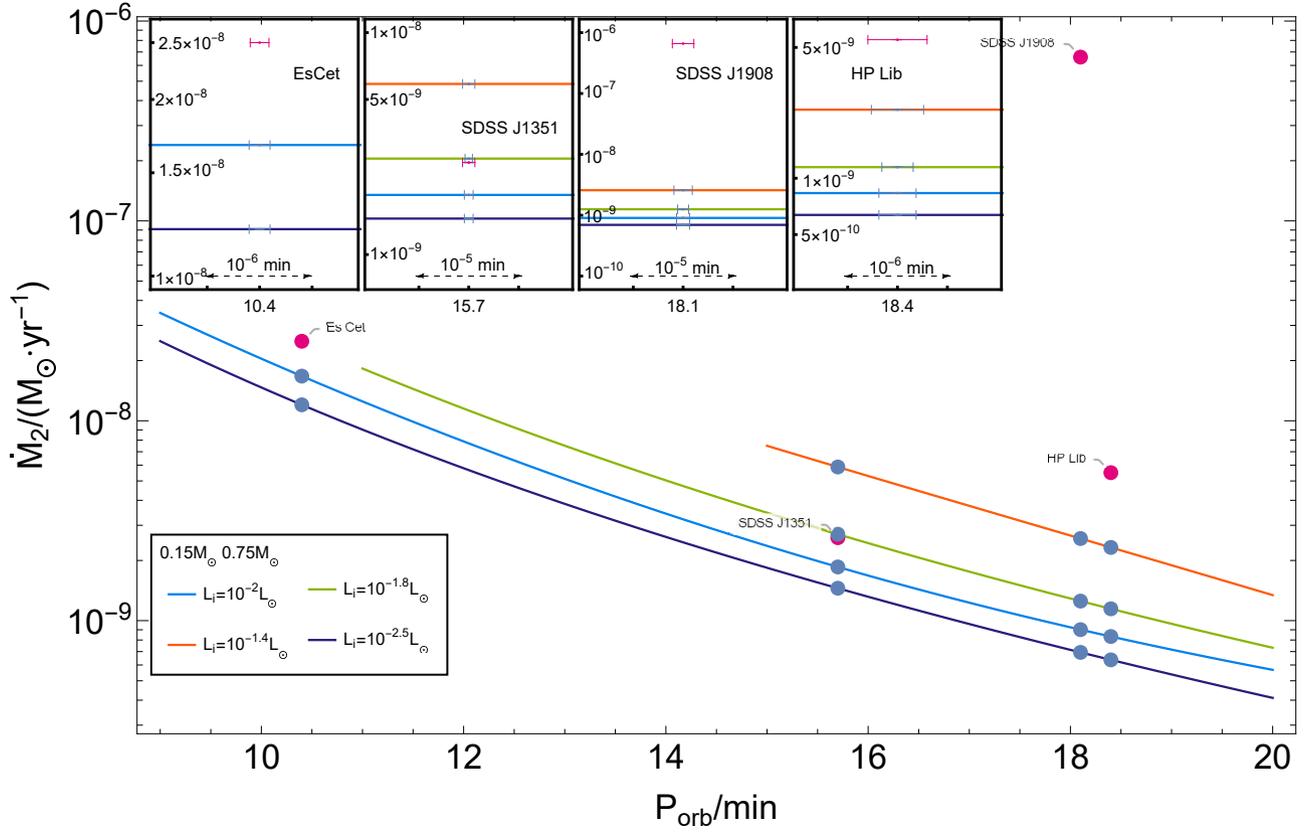}
\caption{GW measurement errors of the MTR and the period estimated by FIM based on LISA for ES Cet, SDSS J1351-0643, SDSS J1908+3940 and HP Lib. We use orbital period of these sources and the corresponding mass transfer rate got by our \texttt{MESA} simulations to make the parameter estimation. We have also zoomed in the regions of measurement at scales of  $O(10^{-6})$ and $O(10^{-7})$ minute to illustrate the GW measurement accuracy.}\label{gw}
\end{figure*}

After the mass transfer begins, $\dot{M}_\mathrm{donor}$ will peak within $10^6$ yrs. More degenerate donors have higher peak $\dot{M}$ and lower $\dot{M}$ at a given age due to the smaller radii (\cite{kaplan2012}). Comparing to a He star, a WD is more degenerate. And thus the two tracks in the $P_{\mathrm{orb}}-\dot{M}_{\mathrm{donor}}$ plane are different. We evolve double white dwarfs with masses $M_1=0.75M_\odot$, $M_2=0.15M_\odot$, using \texttt{binary} of Modules for Experiments in Stellar Astrophysics (\texttt{MESA}, v12778; \cite{MESA1,MESA2,MESA3,MESA4,MESA5}). To construct a He WD and a C/O WD, we evolve a 1.5$M_\odot$ and a 4$M_\odot$ zero-age main-sequence star respectively with metallicity of $Z = 0.02$ until He cores reach to the required masses, $0.15M_\odot$ and $0.75M_\odot$ respectively. And then the hydrogen envelope is stripped with high mass loss rate. Eventually, we evolve the bare He core to $\log(L/L_\odot)=-2$ (\cite{zhangxf2018}) alongside the WD cooling track, avoiding helium ignition. In order to investigate stars with different degeneracy, we let the cooling tracks evolve to $\log(L/L_\odot)=-1.4,~-1.8,~-2.5$ as initial luminosity. As shown in Eqn. \eqref{dftotal}, GWR always leads the orbital frequency to increase and the mass transfer term is the opposite. We focus on the stage where the orbital period is increasing, corresponding to the negative time derivative of the frequency, $\dot{f}<0$. In this phase we are sure that the mass transfer effect has been involved in the binary evolution and then could be measured. In the WD cooling track, materials of the He star with higher luminosity are less compact and the radius is larger. Therefore the donor enters mass transfer at earlier stage with a longer period. Then we analyze the GW measurement of these sources. Fig.~\ref{gw} presents the results of GW measurement errors of the MTR and the period given by FIM based on LISA. Since there have been no GW data yet, we use the photometric results and \texttt{MESA} simulations to set the system parameters and other required orbital parameters are given by \cite{ramsay2018}. In Table \ref{tab:error}, we show the GW measurement uncertainties of the four sources' periods and mass transfer rates. We also show optical measurement errors which are obtained from Gaia DR2 (\cite{ramsay2018}). Results indicate that GW measurement accuracy could improve more than four orders of magnitude comparing to optical observations for those sources which have relatively large mass transfer rate, such as Es Cet. Even for distant sources with small mass transfer rate, the improvement could also reach more than two orders of magnitude. GW detections could measure both mass transfer rate and frequency of a certain white dwarf binary at a level of high accuracy and especially the errors of period are $O(10^{-7})$ minute. Hence future GW measurements would make a significant contribution to improve resolutions in the $P_{\mathrm{orb}}-\dot{M}_{\mathrm{donor}}$ plane. According to such information we can easily distinguish the formation channel of AM CVns.

\begin{table*}
\centering
\caption{GW measurement uncertainties of orbital periods and mass transfer rates. The values and optical measurement errors of the two parameters are from Gaia DR2 (\protect\cite{ramsay2018}).}\label{tab:error}
\begin{tabular}{ccccccc}
\hline
Source & Distance & $P_{\rm orb}$ & $\Delta P_{\rm orb,~GW}$ & $\dot{M}_2$ & $\Delta\dot{M}_{2,~\rm GW}$ & $\Delta\dot{M}_{2,~\rm Gaia~DR2}$\\
& (pc) & (mins) & & $(M_\odot\rm yr^{-1})$ & &\\
\hline
Es Cet & 1584 & 10.4 & $8.9\times10^{-8}$ & $2.5\times10^{-8}$ & $2.0\times10^{-12}$ & $1.6\times10^{-8}$\\
SDSSJ1351 & 1317 & 15.7 & $6.2\times10^{-7}$ & $2.6\times10^{-9}$ & $4.9\times10^{-12}$ & $2.4\times10^{-9}$\\
SDSSJ1908 & 1044 & 18.1 & $1.1\times10^{-6}$ & $6.6\times10^{-7}$ & $5.9\times10^{-12}$ & $3.1\times10^{-7}$\\
HP Lib & 276 & 18.4 & $3.0\times10^{-7}$ & $5.5\times10^{-9}$ & $1.7\times10^{-12}$ & $1.9\times10^{-9}$\\
\hline
\end{tabular}
\end{table*}

\section{Conclusions}\label{sec5}
In this work, we perform a FIM analysis to investigate the measurement of mass transfer rate in WD channel of AM CVn systems using space based GW detectors such as LISA. We assume the loss of orbital angular momentum is only due to the gravitational radiation and introduce mass transfer rate into time derivative of orbital frequency. For a typical AM CVn system with masses of $0.15M_\odot-0.75M_\odot$ and frequency of $5\times10^{-3}$ Hz, we found that GW detection of mass transfer rate could extend the distance to $10^5$ pc (SNR $>5$). For the known AM CVn systems with distances about kpc, the mass transfer rate could be measured as accurate as $5\times10^{-13}M_\odot$/yr. For Milky Way mass-transfering DWDs with SNR $>10$, based on LISA observation, the mass transfer rate could be measured with accuracy $6\times10^{-12}M_\odot$/yr. The numbers of such LISA resolvable semi-detached binaries estimated by most binary population synthesis are several thousands, even though the numbers vary with different binary evolution models (\cite{nelemans2001c,kremer2017,breivik2018}). The GW measurement could cover the early mass transfer phases and the stage $\dot{M}_{\mathrm{donor}}$ peaks. Moreover, surveys by GW could be carried out for sources within 100 kpc. And consequent joint observations with electromagnetic band can be done. The survey could also test accretion theories based on the observation of Milky Way known sources. Such observation can also enlarge the mass transfering binary population.

The three possible channels of AM CVn system formation differ largely in donor star's property especially the donor's mass-radius relationship. we evolve double white dwarfs of $0.15M_\odot-0.75M_\odot$ using \texttt{MESA}. Initially we cool a bare He core to $\log(L/L_\odot)=-2$ as a standard WD following \cite{zhangxf2018}'s procedure. We investigated several luminosities in order to investigate different material compactness. \cite{ramsay2018} have extracted 15 AM CVn stars' mass transfer rate from Gaia DR2 dataset and found they are greater than $P_{\rm orb}-\dot{M}_{\rm donor}$ tracks of a hot donor star predicted by \cite{bildsten2006}. We analyzed four of them which are within LISA verification binaries. Using these optical observations and evolution tracks as input values for FIM, we found that the GW measurement could reduce the error bars of mass transfer rate by at least one order of magnitude and the error bars of orbital period to $O(10^{-7})$ minute. Improvements of the ``resolution'' in the $P_{\mathrm{orb}}-\dot{M}_{\mathrm{donor}}$ plane would significantly help recognize the donor's physical property and exclude evolution models. In current observations, detection for white dwarfs especially for mass transfer rate requires photometric information in a wide range spectrum extending to ultraviolet band (\cite{ramsay2005,ramsay2006}). However, cool white dwarfs with low luminosity are difficult to be measured. Detection by GW performs as a new means to discover cool donors with low luminosity during mass transfer, independent of photometric information and may thereby overcome the limitation of optical observations in the near future.

The discussions are mainly based on increasing orbital period stage ($\dot{f}<0$) and we have possibility of mistaking mass transfering double white dwarfs for detached systems when $\dot{f}>0$. To some extent, we could recognize mass transfer effects in a maximum run time of ten years. The differences between semi-detached and detached binaries could be described by Eqn.~\eqref{dftotal} and the second term would be removed in detached systems. Two sets observation of $f$ and $\dot{f}$ are required: the first year observation result (denoted with subscript of ``$i$'') and the last year observation result (denoted with subscripts of ``$l$''). We could evolve the first term in Eqn.~\eqref{dftotal} with initial values of $f_i$ and $\dot{f}_i$ for ten years and obtain evolved final values $f'_l$. Imposing the observed values to FIM, measurement errors $\Delta f_i$ and $\Delta f_l$ are obtained. If $\mid f'_l-f_l\mid>\mid\Delta f_i+\Delta f_l\mid$, we are still sure that the binary is dominated by some astro-effects, such as mass transfer and L-S coupling, since the frequency and its time derivative could be measured accurately by LISA (\cite{shah2012}). $\dot{f}>0$ represents early mass transfer stage where the mass transfer rate history remains a large amount of uncertainties so far. Extracting mass transfer rate of this phase and reconstructing the evolution tracks are interesting when the GW detection is available.

\section*{Acknowledgments}
We would like to thank Prof. Lev Yungelson for his comments on our manuscript and thank Doc. Yacheng Kang for his correction of formulae. This work was supported in part by the National Key Research and Development Program of China Grant No. 2021YFC2203001 and in part by the NSFC (No.~11920101003, No.~12021003 and No.~12005016). Z. Cao was supported by ``the Interdiscipline Research Funds of Beijing Normal University" and CAS Project for Young Scientists in Basic Research YSBR-006.
\section*{Data Availability}
The data underlying this article will be shared on reasonable request to the corresponding author.

\bibliographystyle{mnras}
\bibliography{ref}

\appendix

\section{Variance-covariance matrix and numerical stability}
In this appendix, we discuss the numerical stability of the FIM analysis involved in the main text. We firstly list the full variance-covariance matrix (VCM) in Table.~\ref{tab:vcm} for the measurement of a set of semi-detached white dwarf binary parameters. The system is a typical AM CVn system with their representative parameter values among this population. The first row lists six parameters, with their respective values provided in the second row. The diagonal elements are absolute uncertainties of each parameter and off-diagonal elements are normalized correlations. As shown in Table.~\ref{tab:vcm}, strong correlations are marked in bold. 

\begin{figure*}
\centering
\includegraphics[width=0.4\textwidth]{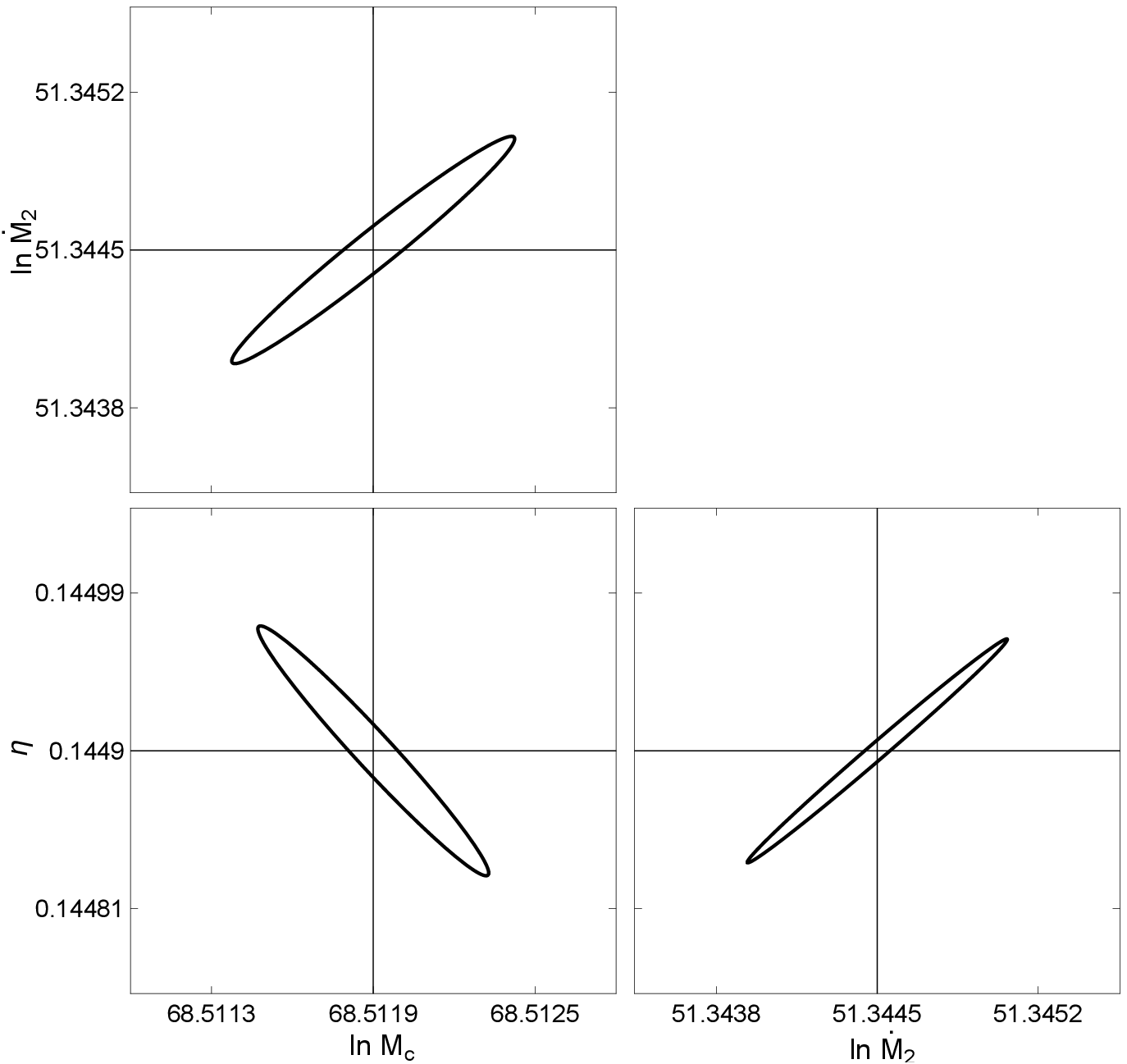}
\caption{Two-dimensional error ellipses of $\ln\dot{M}_2,~\ln M_c,~q$, for a typical AM CVn system. The black solid lines represent $1-\sigma$ ellipses.}\label{ellipse}
\end{figure*}

The results of VCM reveals that the set of parameters ($\dot{M}_2,~M_c,~q$) which define the frequency derivative are highly correlated or anti-correlated. The reason for this is that the frequency changes are limited within a runtime of observation and the signals appear to be similar alongside the gradients of these three parameters. In order to demonstrate reliablility of measurements, two-dimensional $1~\sigma$ error ellipses are shown in Fig.\ref{ellipse}. The elongated error ellipses indicate a relatively high degree of parameter degeneracy. However, it is worth noting that the ellipses imply that there are still discernible variations and constraints that can be used to make measurements of the parameters.

\begin{figure*}
    \centering
    \begin{multicols}{2}
         \subcaptionbox{\label{fig:a}}{\includegraphics[width=0.8\linewidth]{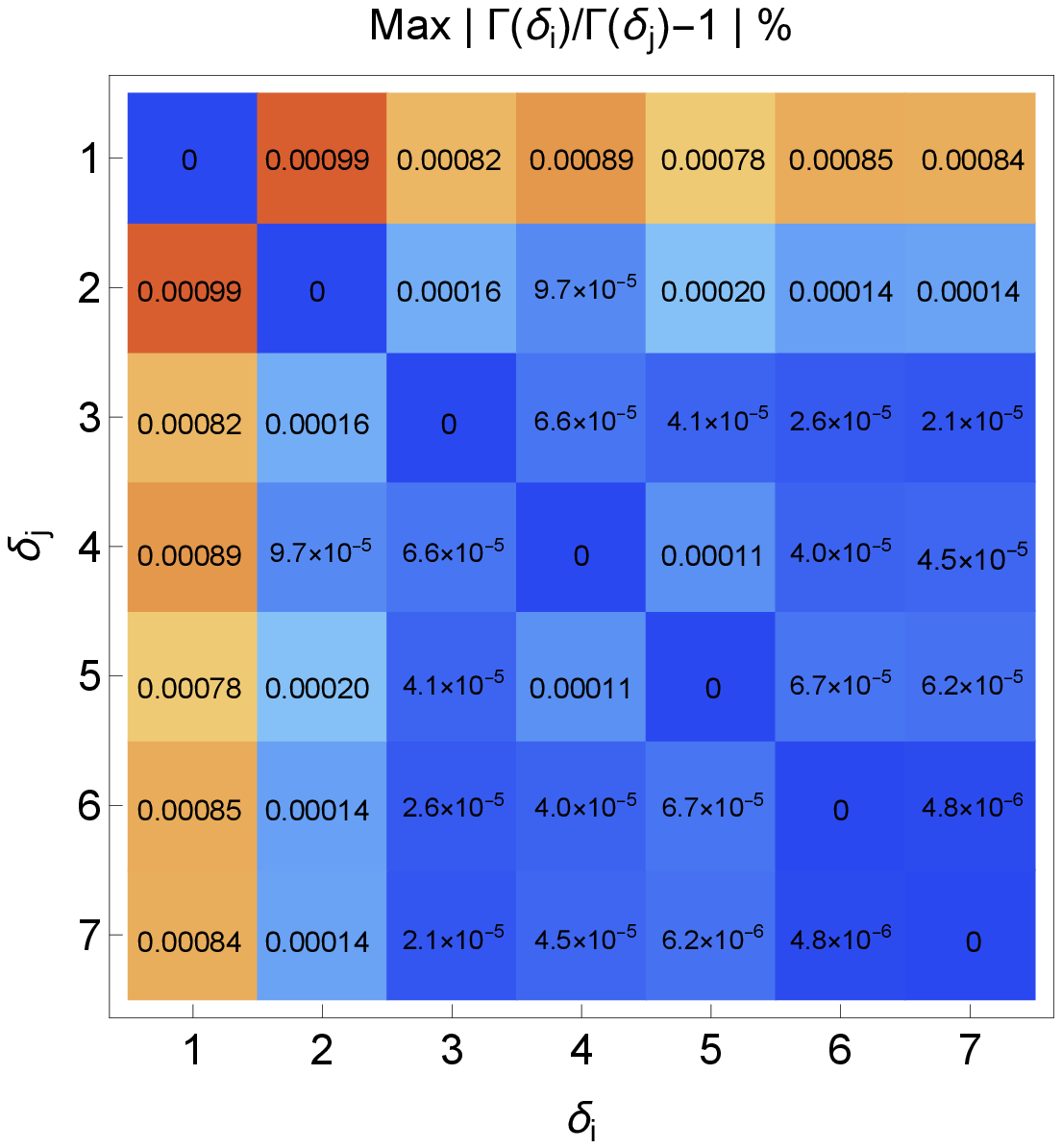}}\par\hspace{-3cm} 
        \subcaptionbox{\label{fig:b}}{\includegraphics[width=0.8\linewidth]{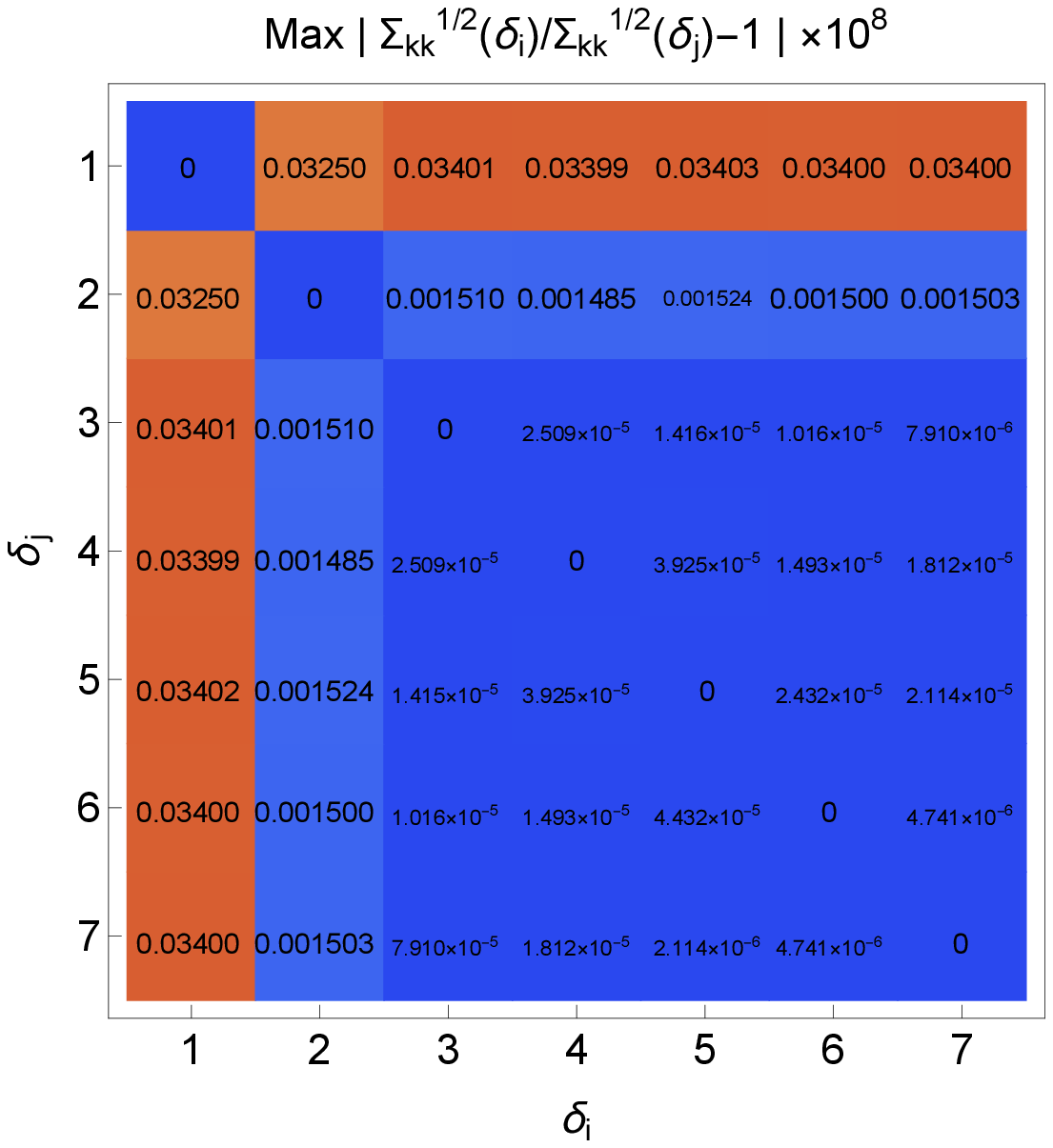}}\par    
   \end{multicols}
    \caption{Maximum relative errors between Fisher matrices computed by different working precisions of integration (left) and maximum relative errors between square root of the diagonal components of the covariance matrices (right).}\label{numstab}    
\end{figure*}

Stabilization of the variance covariance matrix could usually be checked by whether the results are consistent when using different levels of accuracy in numerical derivatives (\cite{shah2012}). In this work, we carried out the analytical expression of derivatives and thereby avoid numerical uncertainties in derivatives. Typically, an unstable FIM implies that small changes in the values of parameters can result in large changes in the estimated parameters and leads to large uncertainties in one or more parameters. Here, numerical uncertainties could only come from numerical integrations. Since the selected integration method is `GlobalAdaptive' in MATHEMATICA, numerical errors of integrations are influenced by working precisions. Therefore, working precisions are varied by 7 levels corresponding to the different numerical errors for integrations in FIM, where the average relative errors $\delta_{i=1,\dots7}= (2.78\times10^{-6},~1.62\times10^{-7},~1.36\times10^{-7},~9.93\times10^{-8},~3.79\times10^{-8},~3.66\times10^{-8},~5.80\times10^{-9})$. Here we use \cite{maselli2022}'s method to check numerical stabilization. Figure \ref{numstab}\,(a) shows the maximum relative error, $|\Gamma_{kl}(\delta_i)/\Gamma_{kl}(\delta_j)-1|\times 100$, computed by using different working precisions of integrations. Results indicate that the Fisher matrices computed by different working precisions differ less than $10^{-5}$. Since it's well known that if the parameters are highly correlated, the adding parameters would dilute the available information and computing the inverse of the Fisher matrix could be numerically difficult. Therefore we apply a truncated singular value decomposition (SVD) approach on $\Gamma$ (\cite{berti2005,pai2013}) and we decompose the Fisher matrix $\Gamma$ as
\begin{equation}
\Gamma=USV^T,\label{SVD}
\end{equation}
where $U,~V$ are orthogonal and $S$ is diagonal, $S={\rm diag}(s_1,~s_2,\dots s_n)$, with sigular values $s_i\geq0$. And then the inverse of $\Gamma$ is given by
\begin{equation}
\Gamma^{-1}=VS^{-1}U^T.
\end{equation}
In this way, we replace the $1/s_i$ by zero when $s_i$ approximates to zero in Eq.~(\ref{SVD}) and eventually obtain a ``pseudoinverse'' of the Fisher matrix which is close to the ``real'' inverse. Figure \ref{numstab}\,(b) shows the maximum relative error between the square root of the diagnoal components of the covariance matrices. Results imply SVD approach helps stabilize the correlation coefficients with differences smaller than $10^{-9}$.

\bsp	
\begin{table*}
\centering
\caption{A typical AM CVn systems with SNR$\sim$321.41.}\label{tab:vcm}
\begin{tabular}{ccccccc}
\hline
 & $ A$ & $f$ & $\dot{M}_2$ & $ M_c$ & $\eta$ & $\phi_0$\\
 & $5.746\times10^{-23}$ & $2\times10^{-3}$Hz & $10^{-8}M_\odot/\mathrm{yr}$ & $0.2753M_\odot$ & 0.1449 &0\\
\hline
$ A$ & $3.209\times10^{-25}$ & -0.1075 & -0.1202 & $9.243\times10^{-2}$ & 0.1202 & $8.272\times10^{-2}$\\
$f$ &  & $3.889\times10^{-11}$ & \bf0.9682 & \bf-0.9681 & \bf-0.9682 & -0.8660 \\
$\dot{M}_2$ & & & $4.319\times10^{-12}$ & \bf-0.9996 & \bf-0.9999 & -0.7453 \\
$ M_c$ & & & & $1.024\times10^{-4}$ & \bf0.9996 & 0.7452\\
$\eta$ & & & & & $5.551\times10^{-5}$ & 0.7453\\
$\phi_0$ & & & & & & $8.341\times10^{-3}$\\
\hline
\end{tabular}
\end{table*}

\label{lastpage}
\end{document}